\documentclass[a4paper,11pt]{article}
\usepackage[a4paper, total={6.5in, 7.5in}]{geometry}
\usepackage{authblk}
\usepackage{amsmath}
\usepackage{amsfonts}
\usepackage{setspace}
\usepackage{multicol}
\usepackage{graphicx}
\usepackage{subcaption}
\usepackage{caption}
\begin{document}
\title{Fluctuations and non-Gaussianity in de Sitter spacetime from holographic entanglement entropy}
\author[a]{Hadyan Luthfan Prihadi\footnote{hadyanluthfanp@s.itb.ac.id}}
\author[a,b]{Freddy Permana Zen\footnote{fpzen@fi.itb.ac.id}}
\author[a]{Donny Dwiputra\footnote{donny.dwiputra@s.itb.ac.id}}
\affil[a]{Theoretical Physics Laboratory, Department of Physics, Institut Teknologi Bandung, Jl. Ganesha 10 Bandung, Indonesia.}
\affil[b]{Indonesia Center for Theoretical and Mathematical Physics (ICTMP), Institut Teknologi Bandung, Jl. Ganesha 10 Bandung,
	40132, Indonesia.}
\maketitle
\abstract{Using entanglement in dS/CFT correspondence, we show that a 4-dimensional de Sitter spacetime in the bulk fluctuates. We calculated the fluctuations of the entanglement entropy, which backreacts with the geometry of the bulk spacetime indicated by the fluctuations of the horizon radius. The variance of the fluctuations is suppressed by $G_N$ and multiplied by $H_{dS}^2$.  Therefore, it will be more significant during the primordial era where it is also well described by de Sitter geometry. We also show that the distribution of the fluctuations is also skewed with the local non-linearity parameter is in the order of $f_{NL}\sim\mathcal{O}(1)$ and we present the probability density function with corrections up to $\sigma_\Phi^3$. }
\newpage
\tableofcontents
\section{Introduction}
Being one of the fundamental forces in nature, gravity should be described by a quantum theory, equivalent to the other known forces. However, applying quantum theory to gravity is not a trivial problem and up until the present, it remains a mystery that needs to be solved.  To put an effort, we might investigate several hints on how a quantum theory of gravity should be, based on known physics. Those clues can give us signatures to quantum gravity which might be observed. There are several works recently that looks for the signs, e.g. signatures in interactions between matter and quantum gravity \cite{Howl2021Signature}, spacetime fluctuations from holographic theories that are detectable in an interferometer \cite{Verlinde2019, Verlinde_2020}, and many more (see, for example, \cite{PhysRevLett.119.240402, Haine2021,Agullo2021PotentialGrav}).\\
\indent In more than two decades, holographic theories, such as the anti-de Sitter/conformal field theory (AdS/CFT) correspondence \cite{Maldacena1997_LargeN,GUBSER1998105,Witten1998AdSCFT}, successfully provide us with significant descriptions on how a quantum theory of gravity should be.  A curved spacetime representing gravity in the bulk might be reconstructed from a CFT with one dimension less living in its boundary \cite{deHaro2001Reconstruction}. Moreover, this is related to the entanglement structure of the CFT. Entanglement entropy of a certain subsystem in the CFT can be calculated by an extremal surface which penetrates to the bulk spacetime \cite{RyuTakayanagi1, RyuTakayanagi2,Hubeny_2007,LewkowyczMaldacena2013}. As a consequence, it is hypothesized that  spacetime in the bulk might be reconstructed from entanglement structure of the CFT, called the entanglement wedge reconstruction \cite{Dong2016WedgeReconstruction}.Since there is no classical counterpart to quantum entanglement, it will be the signature of quantum theory of gravity as well. We may scrutinize possible consequences on a quantum theory of gravity that emerges from entanglement structure of the CFT in its boundary and wish that such signs can be detected at a macroscopic level.\\
\indent Being a thermodynamic quantity, entanglement entropy also obey the first law of thermodynamics, $TdS=dE$, relating a change in entropy and energy (matter) \cite{AlishahihaEntanglementThermodynamics, Bhattacharya2013Thermodynamical,Wong_2013, Kastor2014ChemicalPotential, Rosso2020, Nadi2019} (see also \cite{SwingleUniversalCrossovers} for the relation between entanglement entropy and thermal entropy). Using this knowledge, one can derive the (linearized) Einstein equation---which also relates geometry and matter---from entanglement thermodynamics; this has been done several times until recently \cite{Raamsdonk2010,Faulkner2014, Lashkari2014, Hasegawa2019}.  The idea of deriving Einstein equation from thermodynamics begin with the work of \cite{Jacobson1995} and then renewed by \cite{Padmanabhan_2010, Verlinde2011}. These works give physicists an insight that gravity in our universe might also emerges from quantum information. The long-range contribution to the entanglement entropy in a certain regime behaves like a modified gravity theory that matches the observation of the dark energy and dark matter \cite{Verlinde_2017EmergentGravity}.\\
\indent We study spacetime fluctuations as a consequence of quantum gravity in de Sitter spacetime.  The holographic entanglement entropy of the CFT fluctuates, following the discoveries that entanglement entropy has fluctuations and, for some cases, it can be large \cite{Yurishchev2009Fluctuations,Yurishchev2010Ising} (See also \cite{deBoer2019Capacity}). In the holographic dictionary \cite{JafferisRelativeEntropy2016}, it renders gravitational perturbation in the bulk spacetime, through the entanglement thermodynamics explained previously. It was shown earlier \cite{Verlinde2019, Verlinde_2020} that spacetime fluctuations should be present in this quantum theory of gravity. The aforementioned works proposed that such fluctuations is detectable at a macroscopic distance by using a gravitational waves detector sized interferometer. This phenomenon should be present in quantum theory of gravity in de Sitter spacetime as well, as we show in this paper.\\
\indent Hypothetically, the quantum effects of gravity should be more significant in extreme conditions such as at a strongly curved spacetime near a black hole singularity or during the beginning of the Universe. Therefore, we should aim for the detection of fluctuations of de Sitter spacetime earlier in time. The earliest imprints of the dawn of the Universe that can be detected with electromagnetic radiation so far is the Cosmic Microwave Background (CMB) \cite{PLANCK2018IX}. Temperature fluctuations detected in the CMB is hypothesized to be the quantum fluctuations of the inflaton field, which is a quantum scalar field driving the accelerated expansion of the primordial universe \cite{SenatoreInflation}.  Being an accelerated space-time, inflation era can also be described by a de Sitter spacetime and hence should have the fluctuations from entanglement as well, as we mentioned earlier. In this work, we provide calculations of the spacetime perturbations coming from entanglement fluctuations as the curvature or gravitational perturbation.  We study the statistical behavior of such fluctuations.\\
\indent The spacetime fluctuations from holographic entanglement entropy will compete with the ones coming from the inflaton field itself, especially at the Gaussian level. Therefore, looking for the Gaussian signals in the CMB will be difficult. However, we could go beyond Gaussian level since non-Gaussianity is also commonly investigated in the CMB \cite{PLANCK2018IX,Bartolo2004NonGaussianInflation}. We show that non-Gaussian fluctuations may characterizes the fluctuations from holographic entanglement entropy and separates them with the fluctuations from inflation field which is highly Gaussian \cite{Maldacena_2003}.  This is so because the von Neumann entropy is skewed \cite{Wei_2020Skewness}. If such signal is present in the CMB, then this will become a sign of quantum gravity and entanglement during inflation. Implications of quantum entanglement or the power spectrum during inflation has been studied earlier in \cite{Kanno_2015}, while the non-Gaussianity was investigated in \cite{Bolis2019NonGaussianity}. \\
\indent In this work, we focus on de Sitter spacetime as a curved spacetime which emerges from entanglement as well. de Sitter spacetime is a maximally symmetric solution to the Einstein equation with a constant positive curvature. It represents an accelerated expansion of the Universe and hence can well describe both the dark energy dominated late-time era and the primordial era of inflation.  It has a cosmological horizon and one may associate its area with a gravitational entropy called the Gibbons-Hawking entropy \cite{GibbonsHawking_1977} which obeys the laws of thermodynamics analogous to the black hole thermodynamics \cite{Bekenstein_1974}. In the holographic perspective,  i.e. the dS/CFT correspondence, de Sitter spacetime can be thought to be dual to some CFTs in its asymptotic boundaries \cite{Strominger_dSCFT,Maldacena_2003}. Using this fact, one may associate the entanglement entropy in the CFTs with the Gibbons-Hawking entropy. This has been done earlier by \cite{Arias_2020} by showing, with certain assumptions, that the holographic entanglement entropy of the CFT duals indeed reproduces the Gibbons-Hawking entropy of a 4-dimensional de Sitter spacetime.\footnote{However, this does not trivially generalized to a case where the spacetime dimension is other than 4.}\\
\indent The structure of this work is as follows. In section \ref{sec2}, we briefly review the structure of de Sitter spacetime, its thermodynamics, and holography. In section  \ref{sec3}, we show that there are spacetime fluctuations in de Sitter spacetime which emerges holographically from entanglement that are induced by the fluctuations of the cosmological entanglement entropy. We obtained the variance of the fluctuations in this section and show that the distribution is nearly-Gaussian. In section \ref{sec4}, we look for more hints from the non-Gaussian part of the probability distribution function by calculating the non-linearity parameter $f_{NL}$ and provide the probability distribution function (PDF) up to $\sigma_\Phi^3$ corrections. We conclude our work in section \ref{sec5}.
\section{de Sitter Spacetime and Holography}\label{sec2}
\subsection{de Sitter Spacetime and Thermodynamics}
\begin{figure}
\begin{center}
\includegraphics[scale=0.5]{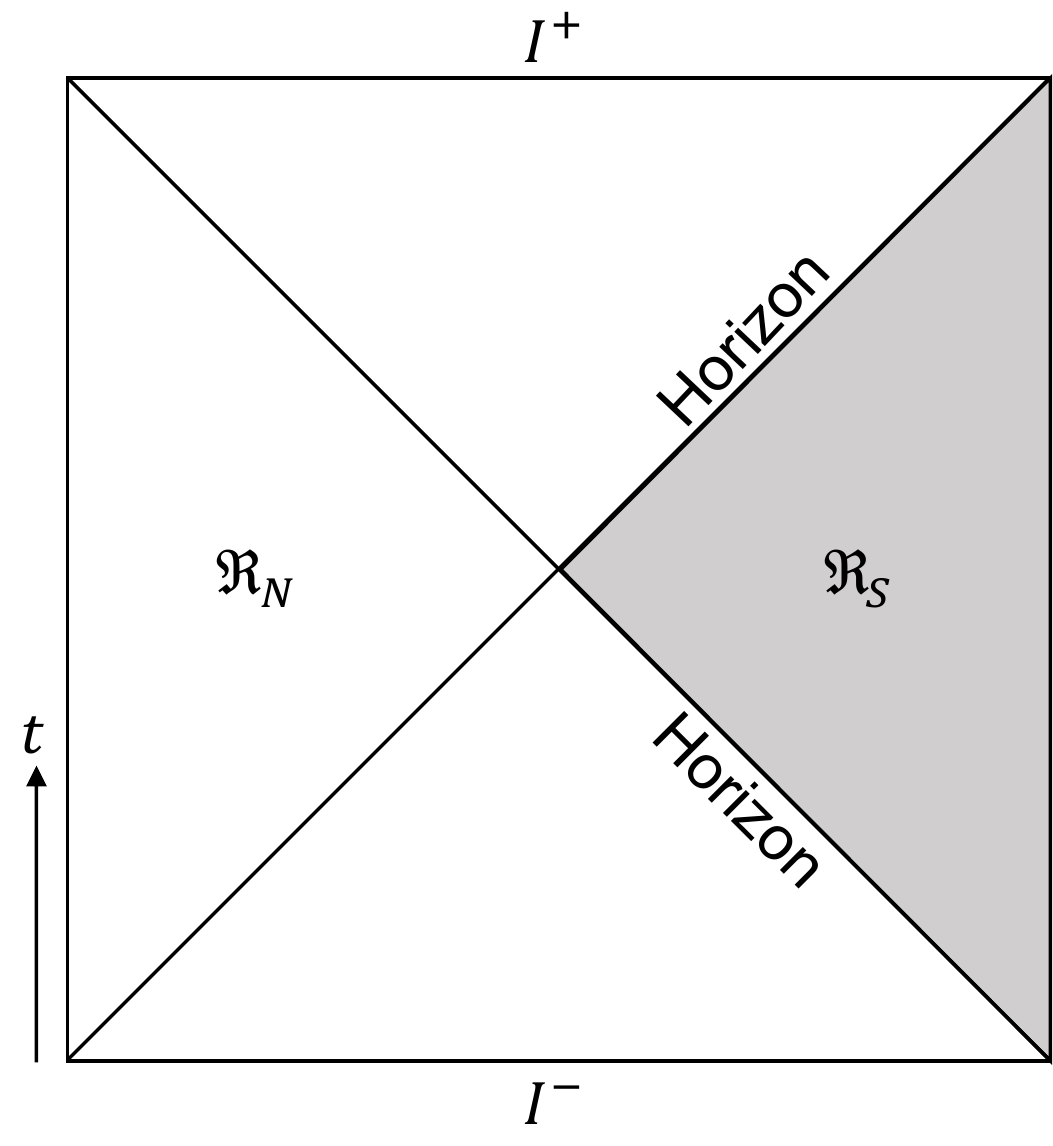}
\caption{Conformal diagram of a global de Sitter spacetime. In 4 dimension, every point in the picture represent a sphere $S^2$. The static patch only covers one region, depicted by the shaded region $\mathfrak{R}_S$. The cosmological horizon is located at $r=\ell_{dS}$, the de Sitter radius. The asymptotic boundaries are located in the infinite future $I^+$ and past $I^-$.}\label{gambardesitter}
\end{center}
\end{figure}

de Sitter (dS) spacetime is defined as a maximally symmetric solution to the Einstein equations. It has a constant positive curvature in oppose to the Anti-de Sitter (AdS) spacetime which has a constant negative curvature. We will begin with a $d+2$-dimensional de Sitter spacetime denoted as $dS_{d+2}$ with a spacetime metric $g_{\mu\nu}(x)$; $\mu,\nu=0,1,...,d+1$. In this work, we will mostly use the static patch of de Sitter spacetime written in the metric form as
\begin{equation}
ds^2=-\bigg(1-\frac{r^2}{\ell_{dS}^2}\bigg)dt^2+\frac{dr^2}{\big(1-\frac{r^2}{\ell_{dS}^2}\big)}+r^2d\Omega_d^2,
\end{equation}
where $r\in[0,\ell_{dS}]$ and $d\Omega_d$ represents a d-sphere. In a full global de Sitter spacetime depicted in figure \ref{gambardesitter}, this coordinate covers only one region, i.e.  a southern Rindler wedge region denoted as $\mathfrak{R}_S$. However, we can have an identical region, with a timelike Killing vector pointing to the opposite direction, denoted as $\mathfrak{R}_N$.\\
\indent Despite being observer-dependent, we are able to associate the cosmological horizon area with a gravitational entropy by using the area law analogous to the Bekenstein-Hawking black hole entropy \cite{Bekenstein_1974, Hawking_1975},
\begin{equation}
S_{dS_{d+2}}=\frac{\ell_{dS}^d\text{Vol}(S^{d})}{4G_N^{(d+2)}},\label{deSitterEntropy}
\end{equation}
where $G_N^{(d+2)}$ is the Newton's gravitational constant in a $d+2$-dimensional spacetime and $\text{Vol}(S^d)$ denotes the volume of a unit sphere $S^d$. When $d=2$, the Gibbons-Hawking entropy \cite{GibbonsHawking_1977}, $S_{GH}$, is recovered with
\begin{equation}
S_{dS_4}=\frac{\pi \ell_{dS}^2}{G_N}=:S_{GH},\label{GibbonsHawking}
\end{equation}
where $G_N$ is now the Newton's gravitational constant in a 4-dimensional spacetime. The de Sitter horizon's temperature may also arise from the periodicity of the Euclidean time in a Euclideanized de Sitter spacetime, giving
\begin{equation}
T_{dS} = \beta^{-1}=\frac{1}{2\pi \ell_{dS}}.\label{deSittertemperature}
\end{equation}
\subsection{AdS/CFT, dS/CFT, and Holographic Entanglement Entropy}
In the last two decades, the AdS/CFT correspondence \cite{Maldacena1997_LargeN} plays a crucial role in studying high energy physics.  It can be applied to calculate the (von Neumann) entanglement entropy of two causally-disconnected regions in conformal field theory (CFT) \cite{RyuTakayanagi1, RyuTakayanagi2}. Consider a (d+1)-dimensional conformal field theory living in the boundary of a (d+2)-dimensional AdS spacetime. Suppose that the boundary CFT is divided into a region inside and outside of a spherical surface, denoted as $B$ and $\bar{B}$ respectively.  Suppose that $a,b=0,1,...,d$ represents indices of the boundary spacetime. Modular energy associated with the region $B$ is given by
\begin{equation}
\hat{H}=\int_B \xi_K^a\hat{T}_{ab}^{\text{CFT}}  dB^b,
\end{equation}
where $\hat{T}_{ab}^{\text{CFT}}$ is the energy-momentum tensor of a CFT in the boundary, $\xi_K^a$ is the Killing vector fields associated with conformal transformations in the boundary, and $dB^b$ is the volume element of the sphere $B$ that points in the time direction. The definition of the modular energy comes from $\hat{H}=-\log\hat{\rho}_B$ and hence expectation value of $\hat{H}$ gives the entanglement entropy which then can be calculated holographically via the Ryu-Takayanagi formula,
\begin{equation}
\langle\hat{H}\rangle=\frac{\text{Area}(\Sigma)}{4G_N}=S_E,
\end{equation}
where $\Sigma$ is the Ryu-Takayanagi surface associated with the region $B$ in the bulk.  This condition represents a vacuum AdS spacetime in the bulk.\\
\indent In a maximally-extended version of Schwarzschild-AdS spacetime, or the eternal black holes in AdS spacetime, one can associate two copies of CFTs, denoted as CFT$_L$ and CFT$_R$, living in two causally-disconnected spatial asymptotics, as its holographic dual \cite{MaldacenaEternal2003}. The full system of the CFT is described by a thermofield double state
\begin{equation}
|\Psi\rangle=\frac{1}{\sqrt{Z_{\text{CFT}}(\beta)}}\sum_n e^{-\frac{\beta E_n}{2}}|n\rangle_L\otimes|n\rangle_R,
\end{equation}
where $Z_{\text{CFT}}(\beta)$ is the partition function and $\beta$ is the inverse temperature associated with the temperature of the Schwarzschild-AdS black hole in the bulk.  This can be seen as a thermal equilibrium between two systems, CFT$_L$ and CFT$_R$, at a temperature $T=\beta^{-1}$. $|n\rangle_L$ and $|n\rangle_R$ are the energy eigenstates of CFT$_L$ and CFT$_R$ respectively.  $|\Psi\rangle$ is an entangled state and the entanglement entropy of one region, i.e. CFT$_L$, given by the reduced density operator $\hat{\rho}_L=\text{Tr}_R(|\Psi\rangle\langle\Psi|)$, reproduces the entropy of the Schwarzschild-AdS black hole in the bulk.  Since the total system is in a pure state, the entanglement entropy of CFT$_R$ is the same. This type of entanglement is similar to the entanglement between left and right region on the Rindler vacuum where the temperature is given by the Unruh temperature \cite{Casini:2011kv} (See also \cite{PhysRevD.96.083531,PhysRevD.103.125005} for more detailed investigations on entanglement between left, right, future and past regions).\\
\indent In de Sitter spacetime, the asymptotic boundaries are in the future and past infinities, denoted as $I^+$ and $I^-$ respectively. It is first hypothesized in \cite{Strominger_dSCFT} that a de Sitter spacetime is dual to two boundary CFTs, denoted as CFT$_+$ and CFT$_-$, living in $I^+$ and $I^-$.  The dual CFTs are Euclidean and might be non-unitary, i.e. the conformal dimension of the associated quantum field might be complex. However, one may write the full system as a thermofield dynamics analogous to the Schwarzschild-AdS case,
\begin{equation}
|\Psi\rangle=\frac{1}{\sqrt{Z_{\text{CFT}}(\beta)}}\sum_n e^{-\frac{\beta E_n}{2}}|n\rangle_{I^+}\otimes|n\rangle_{I^-},\label{deSitterthermofield}
\end{equation}
and discover that the entropy of entanglement between CFT$_+$ and CFT$_-$ reproduces the Gibbons-Hawking entropy in eq.(\ref{GibbonsHawking}) in 4-dimensional de Sitter spacetime. This result has been checked in \cite{Arias_2020} by calculating the entanglement entropy holographically. It also consistent with other papers working on the entanglement entropy in de Sitter spacetime such as \cite{Dong_2018desitterholography,Narayan_2018}. The thermofield prescription in eq. (\ref{deSitterthermofield}) also agrees with the dS/CFT correspondence studied in \cite{Maldacena_2003}. In the later work, the partition function of the CFT is associated with the Hartle-Hawking wavefunction, $\Psi_{dS}=Z_{\text{CFT}}$, and the de Sitter bulk is described by $|\Psi_{dS}|^2$. Hence, there should be two copies of CFTs representing the de Sitter bulk spacetime.\\
\indent There is a work \cite{Sato2015CommentsdSCFT} that calculates the holographic entanglement entropy in dS/CFT from double Wick rotation. The entanglement entropy is given by $S_{dS}\sim(-i)^{d}\ell_{dS}^{d}$ where $\ell_{dS}$ is the radius of a de Sitter spacetime.  It has a multiplication factor of $(-i)^{d}$ compared to the entanglement entropy from AdS/CFT, which is given by $S_{AdS}\sim\ell_{AdS}^{d}$. This means, in a 4-dimensional de Sitter spacetime ($d=2$), we will have an entanglement entropy that is negative-valued. In the next section we argue that the extra negative value comes from the first law of thermodynamics of de Sitter entropy being $TdS=-dE$ instead of $TdS=dE$. This is in agreement with the thermodynamics of de Sitter spacetime presented in \cite{GibbonsHawking_1977}. \\
\indent From the earlier observations of entanglement in de Sitter spacetime, we begin this work by arguing that de Sitter spacetime emerges from entanglement. We study one of the consequences of the argument, which is spacetime perturbations induced by fluctuations of the entanglement entropy.  Studying perturbations in de Sitter spacetime is of particular interest specifically in primordial cosmology with inflationary paradigm since such fluctuations might be able to explain temperature anisotropies observed in the CMB.  Other consequences of de Sitter spacetime being emergent from quantum information is studied in \cite{Verlinde_2017EmergentGravity}.  It is proposed that the emergent gravity in de Sitter spacetime during the late time era may leads to some behaviors that are initially explained by dark energy and dark matter. 
\section{Fluctuations in de Sitter Spacetime from Entanglement}\label{sec3}
\subsection{Fluctuations from Entanglement in AdS/CFT and dS/CFT}
Before we discuss a positively-curved de Sitter spacetime, we would like to mention previous investigation about spacetime fluctuations from entanglement of a finite region in AdS/CFT from recent papers \cite{Verlinde2019, Verlinde_2020}. It is proposed that spacetime fluctuations can be detected at macroscopic distance when we consider a general quantum gravity from holographic principle.  In the context of AdS/CFT, such fluctuations come from a modular Hamiltonian associated to a causal diamond of a finite spatial region in the CFT.  The fluctuations of energy given by the modular Hamiltonian then induce spacetime fluctuations in the AdS bulk (in other word, it "gravitate"). It was also shown that the amount of entanglement, which is determined by the entropy of a Rindler horizon of the causal diamond, determines the magnitude of the fluctuations.\\
\indent Locally, there are slight deviations from the value of $\langle\hat{H}\rangle$, denoted as $\Delta\hat{H}\equiv\hat{H}-\langle\hat{H}\rangle\cdot\hat{\mathbb{I}}$, where $\hat{\mathbb{I}}$ is an identity matrix,  which according to \cite{JafferisRelativeEntropy2016} represents an energy term in the bulk,
\begin{equation}
\Delta\hat{H}=\int_{\mathcal{C}}\xi_H^\mu \hat{T}_{\mu\nu}^{\text{bulk}} d\mathcal{C}^\nu,\label{bulkhamiltonian}
\end{equation}
where $\mathcal{C}$ is the volume of a region in the bulk that is bounded by $\Sigma\cup B$. This means that $\Delta\hat{H}$ sources energy fluctuations in the bulk, and by using Einstein equation, it also generates spacetime perturbations. Since the bulk spacetime is now sourced by a matter term $\langle\hat{T}_{\mu\nu}^{\text{bulk}}\rangle$, the bulk AdS solution is no longer vacuum. The variance of the energy fluctuations from entanglement in AdS/CFT was calculated in \cite{Verlinde_2020}, giving
\begin{equation}
\langle\Delta\hat{H}^2\rangle=\langle\hat{H}^2\rangle-\langle\hat{H}\rangle^2=\frac{\text{Area}(\Sigma)}{4G_N},
\end{equation}
where $\Sigma$ is the area of the Rindler horizon in the bulk. This is exactly the holographic entanglement entropy from AdS/CFT associated with entanglement between the region $B$ and $\bar{B}$. Being dual to the gravitational theory in AdS spacetime using eq. (\ref{bulkhamiltonian}), this energy fluctuations induces metric perturbation written as the variance of the Newton's gravitational potential in the bulk,
\begin{equation}
\langle\Phi^2\rangle=\frac{1}{(d-1)^2}\frac{4G_N}{\text{Area}(\Sigma)}.\label{varianceads}
\end{equation}
The bulk spacetime now exhibits some fluctuations on the top of an empty AdS background.\\
\indent The next question is whether it is possible to extend the calculation in the context of entanglement in dS/CFT to obtain spacetime fluctuations in de Sitter geometry.  In contrary with AdS spacetime, the microscopic description of the de Sitter spacetime is not yet well understood.  However, using certain assumptions presented in the previous section, one may assume that spacetime fluctuations from holographic entanglement entropy can also be present in de Sitter spacetime similar to the mechanism of fluctuations in AdS/CFT. The bulk modular Hamiltonian similar to eq. (\ref{bulkhamiltonian}) arises as the source of spacetime perturbations in de Sitter spacetime.
\subsection{Fluctuations of de Sitter Entropy}
The de Sitter entanglement entropy $S_E(\beta)$ can be calculated from the thermal expectation value of an operator $\hat{S}_E(\beta)=-\log\hat{\rho}(\beta)$, where $\hat{\rho}(\beta)=e^{-\beta\hat{H}}/Z_{dS_4}(\beta)$ is a thermal matrix density operator, i.e. $S_E(\beta)=\langle\hat{S}_E(\beta)\rangle$ with $\langle\cdot\rangle\equiv\text{Tr}(\hat{\rho}(\beta)\cdot)$. The partition function from a Euclidean path integral formalism of a 4-dimensional de Sitter spacetime, $Z_{dS_4}(\beta)$, is equal to the partition function of two copies of the CFT duals, $Z_{\text{CFT}_{\pm}}$. If there are local energy fluctuations, we are able to detect them by observing the variance of the entanglement entropy fluctuations.  The deviation from the average value is defined as $\delta\hat{S}_E(\beta)\equiv\beta\hat{H}$ and the variance can be obtained from
\begin{equation}
\langle(\delta\hat{S}_E(\beta))^2\rangle=\partial_q^2\big(\log\text{Tr}(\hat{\rho}(\beta)^q)\big)\big|_{q=1}=\partial_q^2\big((1-q)S_{\text{R\'en}}^{(q)}(\beta)\big)\big|_{q=1},\label{RMS}
\end{equation}
where $S_{\text{R\'en}}^{(q)}(\beta)$ is the R\'enyi entropy 
\begin{equation}
S_{\text{R\'en}}^{(q)}=\frac{1}{1-q}\log\text{Tr}(\hat{\rho}(\beta)^q),
\end{equation}
satisfying $\lim_{q\rightarrow1}S_{\text{R\'en}}^{(q)}(\beta)=\langle\hat{S}_E(\beta)\rangle$.\\
\indent The R\'enyi entropy can be calculated holographically by calculating the area of a codimension-2 cosmic brane homologous to the entangling region \cite{Dong_2016},
\begin{equation}
\tilde{S}_{\text{mod.}}^{(q)}\equiv q^2\partial_q\bigg(\frac{q-1}{q} S_{\text{R\'en}}^{(q)}\bigg)=\frac{\text{Area}(\text{Cosmic Brane}_q)}{4G_N}.\label{holographicmodular}
\end{equation}
Again the modular entropy $\tilde{S}_{\text{mod.}}^{(q)}(\beta)$ is related to the entanglement entropy $S_E$ by taking the limit of $q\rightarrow1$. In a 4-dimensional de Sitter spacetime, the cosmic brane is a 2-dimensional surface with a tension $\mathcal{T}_q=\frac{1}{4G_N}\big(1-\frac{1}{q}\big)$, as shown in \cite{Arias_2020}. This tension backreacts with the background de Sitter geometry in the bulk. From the definition, the modular entropy is closer to the thermal entropy and hence one can see that it can be obtained by replacing the inverse temperature $\beta$ in the thermal entropy $S_E(\beta)$ with $\beta_q\equiv\beta q$, i.e. $\tilde{S}_{\text{mod.}}^{(q)}(\beta)=S_E(\beta_q)$.  Hence, the variance in eq. (\ref{RMS}) is now given by
\begin{equation}
\langle(\delta\hat{S}_E(\beta))^2\rangle=-\frac{1}{q}\partial_q\tilde{S}_{\text{mod.}}^{(q)}(\beta)\bigg|_{q=1}=-\frac{1}{q}\partial_q S_E(\beta_q)\bigg|_{q=1}.
\end{equation}
One can see that even though we are taking the limit of $q\rightarrow1$ at the end and the cosmic brane becomes tensionless, we are involved with the calculation of the derivative of the cosmic brane with respect to the R\'enyi index $q$. Therefore, we need to perform calculations for each replicated de Sitter geometry that contributes to the variance.\\
\indent The inverse temperature $\beta_q$ and the area of the cosmic brane can be obtained by deforming the de Sitter spacetime due to the existence of a matter in the bulk center.  The static coordinates of $dS_4$ then becomes a 4-dimensional Schwarzschild-de Sitter spacetime
\begin{equation}
ds^2=-\bigg(1-\frac{2G_NM_q}{r}-\frac{r^2}{\ell_{dS}^2}\bigg)dt^2+\frac{dr^2}{\big(1-\frac{2G_NM_q}{r}-\frac{r^2}{\ell_{dS}^2}\big)}+r^2d\Omega^2.
\end{equation}
$M_q$ is the matter in the bulk center and satisfy $M_1=0$, i.e. the cosmic brane becomes tensionless and the original geometry is a pure de Sitter spacetime. In this geometry, the inverse temperature is then given by $T_q\equiv\beta_q^{-1}=-\frac{1}{4\pi}f_q'(r_q)$, where $f_q(r)=1-\frac{2G_NM_q}{r}-\frac{r^2}{\ell_{dS}^2}$, $f_q'(r)\equiv\frac{df_q(r)}{dr}$ and $r_q$ is the horizon radius such that $f_q(r_q)=0$. The area of the cosmic brane then becomes $\text{Area}(\text{Cosmic Brane}_q)=4\pi r_q^2$ and the modular entropy is given by
\begin{equation}
\tilde{S}_{\text{mod.}}^{(q)}(\beta)=S_E(\beta_q)=\frac{\pi r_q^2}{G_N}.\label{modularentropy}
\end{equation}
In the limit of $q\rightarrow 1$, we recover the Gibbons-Hawking entropy in eq. (\ref{GibbonsHawking}) since we have $r_1=\ell_{dS}$.\\
\indent To obtain the variance in eq. (\ref{RMS}), one needs to calculate $\frac{dr_q}{dq}\big|_{q=1}$. This can be achieved by the following. Note that the temperature after the deformation $T_q$ should be $\beta_q^{-1}=(\beta q)^{-1}=\frac{1}{2\pi \ell_{dS}q}$. Therefore, we have another equation for $r_q$, 
\begin{equation}
\frac{2r_q}{\ell_{dS}^2}-\frac{2G_N M_q}{r_q^2}=\frac{2}{\ell_{dS}q}.\label{rqeq}
\end{equation}
On the other hand, $f_q(r_q)=0$ gives us the equation for $M_q$,
\begin{equation}
M_q=\frac{r_q}{2G_N}\bigg(\frac{r_q^2}{\ell_{dS}^2}-1\bigg).
\end{equation}
However, since $0\leq r_q\leq\ell_{dS}$, we always have a negative mass.  This mass is related with the cosmic brane tension $\mathcal{T}_q$ which performs backreacktion to the background geometry and vanishes in the limit of $q\rightarrow1$. Substituting the absolute positive value of $M_q$ into eq. (\ref{rqeq}) and deriving the equation with respect to the R\'enyi index $q$ gives us
\begin{equation}
\frac{dr_q}{dq}=-\frac{2}{q^2}\bigg(\frac{\ell_{dS}r_q^2}{3r_q^2+\ell_{dS}^2}\bigg),\label{rqequation}
\end{equation}
and $\frac{dr_q}{dq} \big|_{q=1}=-\frac{1}{2}\ell_{dS}$. The calculation of the variance in eq. (\ref{RMS}) is then straightforward, giving us
\begin{equation}
\langle(\delta\hat{S}_E(\beta))^2\rangle=-\frac{2}{q}\frac{\pi r_q}{G_N}\frac{dr_q}{dq} \bigg|_{q=1}=\frac{\pi\ell_{dS}^2}{G_N}.\label{variance}
\end{equation}
This is not a surprising result, however. The variance is equal to the entanglement entropy itself, as it is from the calculation of spacetime fluctuations in AdS/CFT \cite{Verlinde_2020}. Nevertheless, the main distinction between the AdS/CFT and dS/CFT formulations lies in the thermodynamic relation.\\
\indent Notice that the temperature of the cosmological horizon in Schwarzschild-de Sitter spacetime is calculated from $T_q=-\frac{f'_q(r_q)}{4\pi}$, which is similar to the usual black hole temperature but with an extra negative sign. We also found the negative term in the calculation of de Sitter entanglement entropy in 4-dimension by performing a double Wick rotation \cite{Sato2015CommentsdSCFT}. Those negatives did not indicate that the entropy itself is negative. It comes from the thermodynamic first law relation of de Sitter geometry \cite{GibbonsHawking_1977} as mentioned in the introduction.  This is so because once we add matter to the de Sitter bulk, we decrease the cosmological horizon's area, i.e. reduces the entropy, rather than increasing it. Now suppose that we want to check whether the fluctuations of de Sitter entanglement entropy indeed induces energy fluctuations in the bulk by treating the mass $M$ as a quantum operator and prove the relation $\beta\hat{M}\propto\delta\hat{S}_E(\beta)$. In order to do that, one have to show that both left and right hand side of the equation obeys a same thermodynamical relation analogous to the first law of black hole thermodynamics. This procedure is similar to the calculation of spacetime fluctuations in AdS/CFT that can be found in section 4 of \cite{Verlinde_2020}.\\
\indent For the left hand side,  one can check directly by performing total differentiation to the function $f(r_H,M)$, where $r_H$ is a horizon radius as the solution of $f(r_H)=0$. This yields to $\beta dM=-dS$ where $dS=\frac{2\pi r_H}{G_N}dr_H$ and it is in agreement with the first law of cosmological horizon in a Schwarzschild-de Sitter spacetime. For the right hand side,  since we have $\delta\hat{S}_E(\beta)=\hat{S}_E(\beta)-S_E(\beta)\cdot\hat{\mathbb{I}}$,  we also have the relation $d\delta\hat{S}_E(\beta)=d\hat{S}_E(\beta)$, without a negative term. Hence, the relation between mass and the entropy fluctuations is given by
\begin{equation}
\beta\hat{M}=-\delta\hat{S}_E(\beta)\quad \rightarrow\quad \delta S_{dS_4}=-2\pi\ell_{dS}M,
\end{equation}
which agree with the result found in \cite{Verlinde_2017EmergentGravity}. The second relation is a direct relation between the change of the de Sitter entropy when we add some mass $M$ to the bulk center. The negative sign of the de Sitter entanglement entropy in 4 dimension is, as mentioned earlier, coming from the thermodynamic relation, i.e. the first law equation relating the change of entropy to the change of energy. This has been realized earlier when studying the thermodynamics of de Sitter spacetime \cite{GibbonsHawking_1977}.
\subsection{Fluctuations of de Sitter Spacetime}
In this section, we correlate the fluctuations of the entanglement entropy of de Sitter spacetime to the metric perturbations.  It is more convenient to work in a spherically-symmetric Friedmann-Robertson-Walker (FRW) patch of a 4-dimensional de Sitter spacetime and perform first order perturbations with respect to this background.  The relation between de Sitter static patch and FRW patch is given by $R=e^{H_{dS}t}\frac{r}{\sqrt{1-H_{dS}^2r^2}}$ and $\tau=t+\frac{1}{2H_{dS}}\ln(1-H_{dS}^2r^2)$, where $H_{dS}=\ell_{dS}^{-1}$ is the Hubble parameter.  The most general scalar perturbations for this spacetime is given by (see, for example \cite{VelezNewtonGauge2018}):
\begin{align}
ds^2=&-(1+2\Phi)d\tau^2+2a(\tau)\partial_R Bd\tau dR+a^2(\tau)\bigg(1+2\Psi+\frac{2}{3}E\bigg)dR^2\nonumber\\
&\;\;\;+a^2(\tau)\bigg(1+2\Psi-\frac{1}{3}E\bigg)R^2d\Omega_2^2,\label{sphericalperturbation}
\end{align}
where $a(\tau)=e^{H_{dS}\tau}$ is the scale factor of a 4-dimensional vacuum de Sitter spacetime. A gauge is chosen such that $B=E=0$ and assuming isotropic perturbations, we have $\Phi=-\Psi$. Hence, we are left with a single scalar $\Phi$ from the metric perturbation and our objective is to correlate it with the entropy fluctuations.\\
\indent In the perturbed metric, the de Sitter horizon's radius is no longer a constant value $\ell_{dS}$ but given by some coordinate-dependent function instead. The radial direction of the perturbed spacetime can be redefined such that the physical radius becomes $\mathcal{R}=(1-\Phi)a(\tau)R$ with the coordinate dependence is encoded in $\Phi$. It is quite difficult to define the cosmological horizon from this perturbed spacetime since there is no time-translation symmetry in the metric and hence there is no Killing vector pointing in the time direction such as $K_\tau K^\tau=-(1-R^2/\ell_{dS}^2)$ in static coordinate. However, we are able to define an apparent horizon $\mathcal{R}_{\text{hor.}}$ such that $(\nabla\mathcal{R}_{\text{hor.}})^2=0$, following the work of \cite{Frolov2003deSitterThermodynamics}.  This new horizon radius then defines the entropy and temperature of a perturbed 4-dimensional de Sitter spacetime,
\begin{equation}
S_{dS_4}=\frac{\pi\mathcal{R}_{\text{hor.}}^2}{G_N}\quad \text{and}\quad T_{dS}=\frac{1}{4\pi\mathcal{R}_{\text{hor.}}}
\end{equation}
\indent The aforementioned work calculated the fluctuations of de Sitter entropy due to metric perturbations induced by the inflaton field during primordial era.  In this work, we use the relation between entropy fluctuations and scalar metric perturbations derived in the previous work, which is given by
\begin{equation}
\delta S_{dS_4}=\frac{2\pi\ell_{dS}^2}{G_N}\Phi,\label{deltaS}
\end{equation}
to say that the entropy fluctuations coming from entanglement fluctuations induces metric perturbation in the de Sitter bulk denoted by $\Phi$ in the same way.  eq. (\ref{deltaS}) was initially obtained from considering the dominating s-wave mode of the fluctuations and hence it only depends on the radius $R$ and $\tau$ of the metric in eq. (\ref{sphericalperturbation}).  It corresponds to large-scale fluctuations in the late-time era of inflation, right before the large scale modes exit the horizon. For angular correlation, we argue that the Green function $G(\vec{R}_1,\vec{R}_2)$ obtained from the two-point correlation function $\langle\Phi(\vec{R}_1)\Phi(\vec{R}_2)\rangle$, when expanded in the spherical harmonics functions such as in \cite{Verlinde2019}, only dominated by the large-scale modes. Hence, in this work, we are working with $\Phi$ as a one-dimensional statistical variable obeying some PDF representing the fluctuations of de Sitter horizon's radius.\\
\indent Since our paradigm is that de Sitter horizon entropy is the entanglement entropy, the variance of the entanglement entropy in eq.  (\ref{variance}) is equal to the variance of de Sitter horizon entropy fluctuations in eq. (\ref{deltaS}). From this fact, we now have the variance of $\Phi$ itself, which is given by
\begin{equation}
\langle\Phi^2\rangle=\frac{G_N}{4\pi}H_{dS}^2.
\end{equation}
This value is proportional to $G_N\simeq10^{-38}\text{ GeV}^{-2}$ with the prefactor $\frac{1}{4\pi}$ arise as the normalization constant from the averaging process. The result also agrees with the calculation of fluctuations of Newton's gravitational potential in AdS/CFT in eq. (\ref{varianceads}). We would like to mention several things about the value of this variance. It is indeed supressed by $G_N$ and hence small. Furthermore, the value also multiplied by $H_{dS}^2$.  Such fluctuations, during the present dark-energy dominated era when the value of the Hubble parameter is around $H_{dS,0}\simeq10^{-42}\text{ GeV}$, is nearly impossible to measure. Nevertheless, this will more likely to be significant during the inflation era when the fluctuations is enhanced by the factor of $H_{dS}\simeq10^{13}\text{ GeV}$. In that timescale, we will have $\langle\Phi^2\rangle\sim\mathcal{O}(10^{-12})$, which is comparable to the power spectrum of metric perturbations generated by the energy fluctuations of the inflaton field slowly rolling the potential during inflation \cite{SenatoreInflation}. If the geometric perturbations of de Sitter spacetime induced by entanglement entropy fluctuations present during the inflation era, then we should be able to observe them in the CMB.\\
\indent The value of $\Phi$ fluctuates around $\langle\Phi\rangle=0$. Those small fluctuations are in the order of $\Phi\sim\sqrt{G_N}H_{dS}$ and the width of the distribution is given by $\sigma_\Phi^2\equiv\langle\Phi^2\rangle$.  The statistics is then well described by a Gaussian normal distribution, i.e. the variance $\sigma_\Phi$ alone completely characterizes the full Gaussian PDF, which is given by
\begin{equation}
P_G(\Phi)=\frac{1}{\sqrt{2\pi\sigma_\Phi^2}}\exp\bigg(-\frac{\Phi^2}{2\sigma_\Phi^2}\bigg).
\end{equation}
However, it is quite challenging to observe such fluctuations since it is some order of magnitudes smaller than the power spectrum generated by the time delay during inflation that has the order of $\mathcal{O}(10^{-10})$, which is then observed in the CMB. Moreover, temperature fluctuations observed in the CMB are also nearly-Gaussian. In other words, we need something to distinguish the signal coming from entanglement fluctuations and other probes such as the fluctuations of the inflaton field. In this case, we go beyond Gaussian distribution and look for some hints there.
\section{Beyond Gaussian Distribution}\label{sec4}
In the previous section, we have already shown that spacetime fluctuations induced by the entanglement entropy fluctuations of de Sitter spacetime is well described by a Gaussian statistics. In this section, we will see whether any small deviation from Gaussian might be present and study its significance. If it exist, then the full one-point (single-variate) PDF for $\Phi$ can be represented by the Edgeworth expansion (see, for example \cite{KofmanPDF1995})
\begin{align}
P(\Phi)=P_G(\Phi)\bigg[1&+\sigma_{\Phi}\frac{S_3^{\Phi}}{6}H_{e_3}(\nu)+\sigma_\Phi^2\bigg(\frac{S_4^{\Phi}}{24}H_{e_4}(\nu)+\frac{(S_3^{\Phi})^2}{72}H_{e_6}(\nu)\bigg)\nonumber\\
&+\sigma_\Phi^3\bigg(\frac{S_5^\Phi}{120}H_{e_5}(\nu)+\frac{S_4^\Phi S_3^\Phi}{144}H_{e_7}(\nu)+\frac{(S_3^\Phi)^3}{1296}H_{e_9}(\nu)\bigg)+...\bigg],\label{edgeworth}
\end{align}
where
\begin{equation}
S_n^\Phi\equiv\frac{\kappa_n^\Phi}{(\kappa_2^\Phi)^{(2n-2)}}=\frac{\kappa_n^\Phi}{(\sigma_\Phi)^{(2n-2)}},\label{sn}
\end{equation}
and $H_{e_n}(\nu)=(-1)^ne^{\nu^2}\frac{d^ne^{-\nu^2}}{d\nu^n}$ are the Hermite polynomials where $\nu\equiv\Phi/\sigma_\Phi$. $\kappa_n^\Phi$ is the n-th cumulant ($n=2,3,4,...$) of the distribution of $\Phi$, which is related to the central moments $\mu_n^\Phi\equiv\langle\Phi^n\rangle$ by
\begin{align*}
\kappa_2^\Phi&=\mu_2^\Phi\;\;\;\text{(variance)},\\
\kappa_3^\Phi&=\mu_3^\Phi\;\;\;\text{(skewness)},\\
\kappa_4^\Phi&=\mu_4^\Phi-3(\mu_2^{\Phi})^2\;\;\;\text{(kurtosis)},\quad&& \mu_4^\Phi=\kappa_4^\Phi+3(\kappa_2^\Phi)^2,\\
\kappa_5^\Phi&=\mu_5^\Phi-10\mu_3^\Phi\mu_2^\Phi,\quad&& \mu_5^\Phi=\kappa_5^\Phi+10\kappa_3^\Phi\kappa_2^\Phi.
\end{align*}
This expansion converges as long as the fluctuations are nearly-Gaussian and $\sigma_\Phi$ is relatively small. In fact, we have $\Phi\sim\sigma_\Phi\sim\mathcal{O}(\sqrt{G_N})$ and hence we then consider the Edgeworth expansion as the expansion in the power of $\sigma_\Phi\sim\mathcal{O}(\sqrt{G_N})$. In this work, we calculate the PDF up to $\sigma_\Phi^3$ correction.\\
\indent In order to calculate $\kappa_n^\Phi$, we first calculate the cumulants of the distribution of $\delta S_E$, $\kappa_n^{S_E}$.  Both $\kappa_n^\Phi$ dan $\kappa_n^{S_E}$ are related by
\begin{equation}
\kappa_n^\Phi=\bigg(\frac{G_N}{2\pi\ell_{dS}^2}\bigg)^n\kappa_n^{S_E},\label{cumulantsrelation}
\end{equation}
from the entropy and curvature perturbation relation in eq. (\ref{deltaS}). Note that in the previous section, we calculate the variance, i.e. the second cumulant, $\langle(\delta\hat{S}_E(\beta))^2\rangle$ from eq. (\ref{variance}). It is not difficult to show that the third cumulant (skewness) and the fourth cumulant (kurtosis) can be calculated from the third and fourth derivative of $(1-q)S_{\text{R\'en}}^{(q)}$ respectively. This will generalize to the n-th cumulant and gives us the relation
\begin{equation}
\kappa_n^{S_E}=(-1)^n\partial_q^n\big(\log\text{Tr}(\hat{\rho}^q)\big)\big|_{q=1}=(-1)^{n+1}\partial_q^{n-2}\bigg(\frac{1}{q}\partial_q\tilde{S}_{\text{mod.}}^{(q)}\bigg)\bigg|_{q=1},\label{cumulants}
\end{equation}
for $n>2$. Again, the calculation of the cumulants involve higher order derivatives of the modular entropy with respect to the R\'enyi index $q$ and hence the backreaction of the cosmic brane tension to each replicated manifold plays a significant role.
\subsection{Skewness of the Entanglement Entropy}
We will begin with the third cumulant or the skewness which is given by
\begin{equation}
\kappa_3^{S_E}=-\partial_q^3\big(\log\text{Tr}(\hat{\rho}^q)\big)\big|_{q=1}=\partial_q\bigg(\frac{1}{q}\partial_q\tilde{S}_{\text{mod.}}^{(q)}\bigg)\bigg|_{q=1}.\label{skewnesscalculation}
\end{equation}
The modular entropy is given by eq. (\ref{modularentropy}) and hence we have
\begin{align}
\kappa_3^{S_E}&=\frac{d}{dq} \bigg(\frac{2\pi r_q}{G_N q}\frac{dr_q}{dq} \bigg)\bigg|_{q=1}\nonumber\\
&=\bigg\{-\frac{1}{q^2}\frac{2\pi r_q}{G_N}\frac{dr_q}{dq} + \frac{2\pi }{G_N q}\bigg(\frac{dr_q}{dq}\bigg)^2+\frac{2\pi r_q}{G_N q}\frac{d^2r_q}{dq^2}\bigg\}\bigg|_{q=1}.
\end{align}
The second derivative of $r_q$ can be calculated directly from eq. (\ref{rqequation}) by deriving it one more time with respect to $q$, giving us
\begin{equation}
\frac{d^2r_q}{dq^2}=\frac{4}{q^3}\bigg(\frac{\ell_{dS}r_q^2}{3r_q^2+\ell_{dS}^2}\bigg)-\frac{2}{q^2}\bigg(\frac{2\ell_{dS} r_q}{3r_q^2+\ell_{dS}^2}\bigg)\frac{dr_q}{dq}+\frac{12}{q^2}\bigg(\frac{\ell_{dS} r_q^3}{(3r_q^2+\ell_{dS}^2)^2}\bigg)\frac{dr_q}{dq}.
\end{equation}
Taking the limit of $q\rightarrow 1$ and using $r_1=\ell_{dS}$, we have $\frac{d^2r_q}{dq^2}\big|_{q=1}=\frac{9}{8}\ell_{dS}.$ Plugging this back to eq.  (\ref{skewnesscalculation}) and using $\frac{dr_q}{dq}\big|_{q=1}=-\frac{1}{2}\ell_{dS}$, we have
\begin{equation}
\kappa_3^{S_E}=\frac{15}{4}\frac{\pi\ell_{dS}^2}{G_N}.
\end{equation}
From the relation in eq. (\ref{cumulantsrelation}), we obtain the skewness of the distribution of $\Phi$ as
\begin{equation}
\kappa_3^\Phi=\frac{15}{32\pi^2}\frac{G_N^2}{\ell_{dS}^4}=\frac{15}{32\pi^2}G_N^2H_{dS}^4.
\end{equation}
\indent This is a rather interesting result. First, we see that the third cumulant of $\Phi$ is suppressed by $G_N^2$. This agrees with the calculations done in \cite{Wei_2020Skewness} which showed that the statistical distribution of the von Neumann entropy is indeed skewed with the skewness is suppressed by the dimension of the Hilbert space $d$ in the limit of $d\rightarrow\infty$. In the $G_N\sim\frac{1}{N^2}$ prescription, where $N$ is the number of degrees of freedom representing the dimension of the CFT Hilbert space, the two results agree precisely, i.e. the skewness is in order of $\mathcal{O}(1/N^4)$ for large-$N$ limit.\\
\indent Second, in spite of being highly suppressed by $G_N^2$, we may calculate the local non-linearity parameter $f_{NL}$, defined from
\begin{equation}
\Phi=\Phi_G+f_{NL}(\Phi_G^2-\langle\Phi_G^2\rangle),\label{nonlinear}
\end{equation}
where $\Phi_G$ is a Gaussian random field, which is not suppressed.  Note that the definition of $f_{NL}$ is compatible with the Edgeworth expansion in eq. (\ref{edgeworth}). From the definition, we may calculate the $f_{NL}$ from
\begin{equation}
f_{NL}\simeq\frac{\langle\Phi^3\rangle}{6\langle\Phi^2\rangle^2}=\frac{\kappa_3^\Phi}{6\sigma_\Phi^4},
\end{equation} 
and obtain $f_{NL}=\frac{5}{4}$.  This observation indicates that the non-Gaussianity, despite being small, might still be detectable since we have $f_{NL}\sim\mathcal{O}(1)$ and not suppressed by any small parameter. In comparison, note that the local non-linearity parameter observed in the CMB by PLANCK 2018 collaboration gives us $f_{NL}=-0.9\pm5.1$ \cite{PLANCK2018IX}, which is still in the order of $\mathcal{O}(1)$. The observation of $f_{NL}\sim\mathcal{O}(1)$ rules out any single-field slow-roll inflation model which have $f_{NL}\sim\mathcal{O}(\varepsilon,\eta)$, where $\varepsilon,\eta$ are the small slow-roll parameters \cite{Maldacena_2003}.
\subsection{Probability Distribution Function up to $\sigma_\Phi^3$}
\begin{figure}
\begin{center}
\hspace{-3.5cm}
\valign{#\cr
  \hbox{%
    \begin{subfigure}{.5\textwidth}
    \includegraphics[scale=0.8]{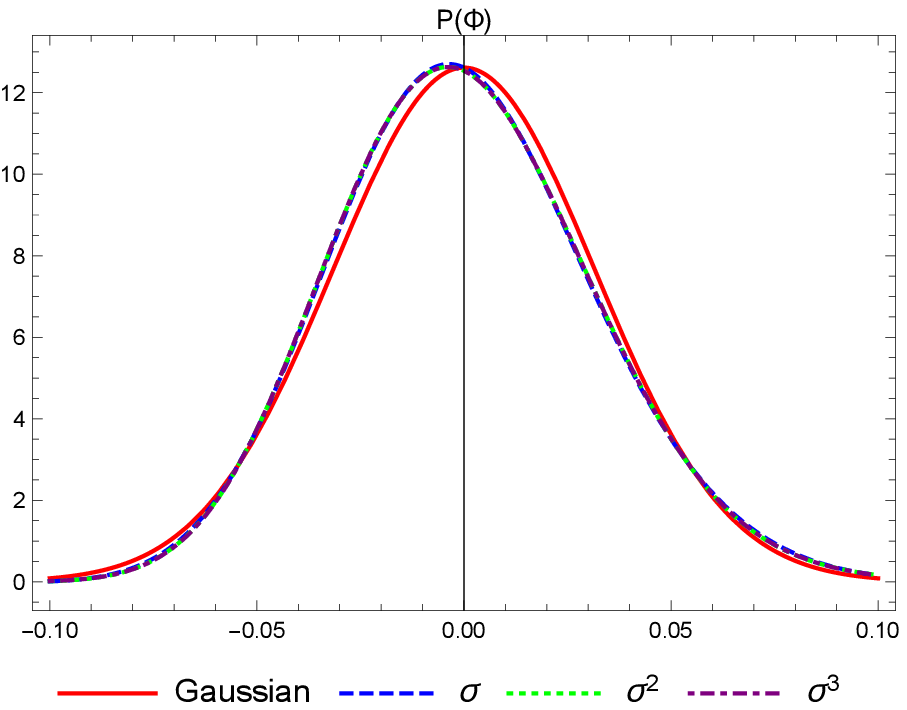}
    \end{subfigure}%
  }\cr
  \hbox{%
    \begin{subfigure}{.0\textwidth}
    \includegraphics[scale=0.65]{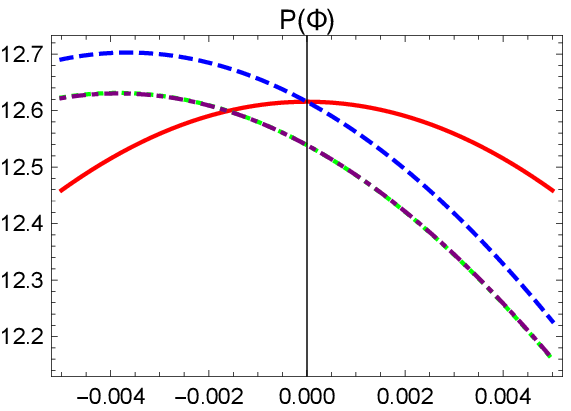}
    \end{subfigure}%
  }
  \hbox{%
    \begin{subfigure}{.0\textwidth}
    \includegraphics[scale=0.67]{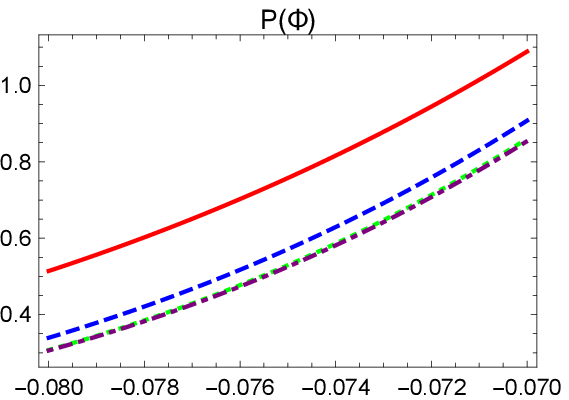}
    \end{subfigure}%
  }\cr
}
\caption{The shape of the PDF with corrections up to $\sigma_\Phi$ (dashed blue line), $\sigma_\Phi^2$ (dotted green line), $\sigma_\Phi^3$ (dot-dashed purple line) compared with the original Gaussian distribution (solid red line). The variance is greatly rescaled to $\sigma_\Phi^2=10^{-3}$ in order to provide a visible deviation. The left image is the plot ranging from $-0.1$ to $0.1$ while the top right and bottom right images represent the zoomed-in parts of the distribution.}\label{figureplot}
\end{center}
\end{figure}

\indent In this part, we present the Edgeworth expansion in eq. (\ref{edgeworth}) up to $\sigma_\Phi^3$. Hence, we need to calculate $\kappa_4^\Phi$ and $\kappa_5^\Phi$. The calculation is straightforward. From eq. (\ref{cumulants}) and the modular entropy in eq. (\ref{modularentropy}) we have
\begin{align}
\kappa_4^{S_E}&=-\frac{d^2}{dq^2} \bigg(\frac{2\pi r_q}{G_N q}\frac{dr_q}{dq}\bigg)\bigg|_{q=1}\nonumber\\
&=-\frac{d}{dq} \bigg\{-\frac{1}{q^2}\frac{2\pi r_q}{G_N}\frac{dr_q}{dq}+\frac{2\pi }{G_N q}\bigg(\frac{dr_q}{dq}\bigg)^2+\frac{2\pi r_q}{G_N q}\frac{d^2r_q}{dq^2}\bigg\}\bigg|_{q=1}.\label{kurtosiscalculation}
\end{align}
Since we already have $\frac{d^2r_q}{dq^2} \big|_{q=1}=\frac{9}{8}\ell_{dS}$, the first and the second terms can be calculated. This gives us
\begin{align}
\frac{d}{dq}\bigg(\frac{1}{q^2}\frac{2\pi r_q}{G_N}\frac{dr_q}{dq}\bigg)\bigg|_{q=1}&=-\frac{2}{q^3}\frac{2\pi r_q}{G_N}\frac{dr_q}{dq}\bigg|_{q=1}+\frac{1}{q^2}\frac{2\pi}{G_N}\bigg(\frac{dr_q}{dq}\bigg)^2\bigg|_{q=1}+\frac{1}{q^2}\frac{2\pi r_q}{G_N}\frac{d^2r_q}{dq^2}\bigg|_{q=1}\nonumber\\
&=\frac{19}{4}\frac{\pi\ell_{dS}^2}{G_N},
\end{align}
and
\begin{align}
-\frac{d}{dq}\bigg(\frac{2\pi}{G_N q}\bigg(\frac{dr_q}{dq}\bigg)^2\bigg)\bigg|_{q=1}&=\frac{2\pi}{G_N q^2}\bigg(\frac{dr_q}{dq}\bigg)^2\bigg|_{q=1}-\frac{4\pi}{G_N q}\bigg(\frac{dr_q}{dq}\bigg)\frac{d^2r_q}{dq^2}\bigg|_{q=1}\nonumber\\
&=\frac{11}{4}\frac{\pi\ell_{dS}^2}{G_N}.
\end{align}
For the third term, we need to calculate $\frac{d^3 r_q}{dq}\big|_{q=1}$. The detailed calculation is provided in the appendix A and it gives $\frac{d^3r_q}{dq}\big|_{q=1}=-\frac{105}{32}\ell_{dS}$. Hence, the value of $\kappa_4^{S_E}$ is given by $\kappa_4^{S_E}=\frac{279}{16}\frac{\pi\ell_{dS}^2}{G_N}$ and the kurtosis of $\Phi$ and $S_4^\Phi$ is given by
\begin{equation}
\kappa_4^\Phi=\frac{279}{256\pi^3}G_N^3 H_{dS}^6,\;\;\;\text{and}\;\;\;S_4^\Phi=\frac{279}{4}.
\end{equation}
\indent The calculation of $\kappa_5^\Phi$ is, as expected, a bit longer.  We calculate $\kappa_5^{S_E}$ from
\begin{equation}
\kappa_5^{S_E}=\frac{d^3}{dq^3}\bigg(\frac{2\pi r_q}{G_Nq}\frac{dr_q}{dq}\bigg)\bigg|_{q=1}.
\end{equation}
More detailed calculations are given in the appendix. We have $\frac{d^4r_q}{dq^4}\big|_{q=1}=\frac{1283}{128}\ell_{dS}$, and hence $\kappa_5^{S_E}=\frac{5957}{64}\frac{\pi\ell_{dS}^2}{G_N}$. The fifth cumulant of $\Phi$ and $S_5^\Phi$ is given by
\begin{equation}
\kappa_5^\Phi=\frac{5957}{2048\pi^4}G_N^4H_{dS}^8,\;\;\;\text{and}\;\;\; S_5^\Phi=\frac{5957}{8}.
\end{equation}
\indent Now we are ready to write the PDF with corrections up to $\sigma_\Phi^3$ and it is given by
\begin{align}
P(\Phi)=P_G(\Phi)\bigg[1&+\sigma_{\Phi}\frac{5}{4} H_{e_3}(\nu)+\sigma_\Phi^2\bigg(\frac{93}{32}H_{e_4}(\nu)+\frac{225}{228}H_{e_6}(\nu)\bigg)\nonumber\\
&+\sigma_\Phi^3\bigg(\frac{5957}{960}H_{e_5}(\nu)+\frac{465}{128}H_{e_7}(\nu)+\frac{125}{384}H_{e_9}(\nu)\bigg)+...\bigg]
\end{align}
If the values of $S_n^\Phi$ are small, i.e. smaller or much smaller than one, then the distribution of $\Phi$ is practically Gaussian. However, as we see from the previous calculations to at least $S_5^\Phi$, the values of $S_n^\Phi$ are larger than one.  Hence, we conclude that the distribution of $\Phi$ deviates from Gaussian distribution and the deviation is small but still visible. The shape of the PDF, with corrections up to $\sigma_\Phi$, $\sigma_\Phi^2$, and $\sigma_\Phi^3$ compared with the original Gaussian distribution is shown in figure (\ref{figureplot}). Calculating the PDF to all orders by finding the higher cumulants is, in principle, possible in order to obtain the full statistical description of $\Phi$.  However, we see that adding more corrections to the PDF higher than $\sigma_\Phi^3$ is of no use (at least for now) since the PDF line will coincide with the one with corrections only up to $\sigma_\Phi^3$.
\section{Conclusions}
\label{sec5}
In this paper, we show that a de Sitter spacetime which emerges from entanglement produces spacetime fluctuations. Using the dS/CFT correspondence as the main paradigm, this shows that such fluctuations should exist in a quantum theory of gravity in de Sitter bulk. Therefore, we suggest that any detection of the fluctuations will be a signature of quantum gravity in de Sitter spacetime.  In a $\Lambda$CDM universe, it is highly likely that our universe undergo two eras that are well described by de Sitter geometry, which are the late-time accelerating expansion during the dark energy domination era and the primordial era during inflation.\\
\indent We calculate the variance of the fluctuations of de Sitter horizon length, encoded in the curvature perturbation $\Phi$ as the scalar part of the metric perturbation. The variance is suppressed by the $G_N$ factor and multiplied by $H_{dS}^2$. For this reason, such fluctuations are more significant during the primordial era, i.e. when the value of $H_{dS}$ is large enough. In the present era, however, the fluctuations are effectively zero and trying to measure them would be impractical. We assume that the fluctuations are nearly-Gaussian and the value of the variance is around $\sigma_\Phi^2\sim\mathcal{O}(10^{-12})$ during the primordial ere, where $H_{dS}\simeq10^{13}$ GeV. This value has the same order of magnitude with spacetime perturbations generated by energy fluctuations of the inflaton field. Unfortunately, it is some order of magnitude smaller (in the order of $\varepsilon$, the slow-roll parameter) than the power spectrum generated by time delay, which is in the order of $\mathcal{O}(10^{-10})$  \cite{SenatoreInflation} and hence observing the Gaussian part of the fluctuations seems not too promising.\\
\indent To obtain a more unique feature of fluctuations in de Sitter spacetime from entanglement entropy, we look for more hints in the non-Gaussian part of the PDF. We found that the local non-linearity parameter defined in eq. (\ref{nonlinear}), calculated from the skewness of the distribution, is given by $f_{NL}\sim\frac{5}{4}$, which is in the order of $\mathcal{O}(1)$. This infers that the distribution of $\Phi$ has some non-linear feature. The value of $f_{NL}$ is large enough, i.e. it is not suppressed by any small parameter such as the slow-roll parameters $\varepsilon$ and $\eta$ in the single-field slow-roll inflation,  to proclaim that there is non-Gaussianity in the distribution of the fluctuations that can be observed. In this work, we also calculate the PDF as an expansion of the full PDF around Gaussian distribution up to $\sigma_\Phi^3$ correction. In principle, the PDF can be calculated to all orders by deriving the equations iteratively. The knowledge of the full PDF to all orders is important to fully understand the behavior of the fluctuations.\\
\indent As a final remark, we suggest that the calculation of the fluctuations in de Sitter spacetime from holographic entanglement entropy might be applied to the inflationary universe. In single-field slow-roll inflation, the accelerating universe is driven by some scalar field $\phi$ called the inflaton. The fields coupled minimally to gravity and hence their quantum fluctuations $\delta\phi$ backreact with the background geometry, generating scalar curvature perturbations to the metric, similar with $\Phi$. In this work, we see $\Phi$ arise from the fluctuations of the entanglement entropy of de Sitter horizon. Thus, during inflation, which also well described by de Sitter spacetime, fluctuations from entanglement might also contribute to the (scalar part of the) spacetime perturbations. Hypothetically, the fluctuations and non-Gaussianity from entanglement might be present during the late time era of inflation, right before the fluctuations exit the horizon.  These are large scale fluctuations and hence it will re-enter the horizon after recombination era. Since it will be constant outside the horizon,  it should also be present in the temperature fluctuations observed in the CMB, similar to the Sachs-Wolfe effect \cite{SachsWolfe1967}. 
\section*{Acknowledgement}
Authors would like to thank Riset ITB and Kementrian Ristekdikbud Indonesia for financial support. H. L. P. also thank Getbogi Hikmawan and other members of theoretical physics group for useful discussions.
\bibliographystyle{JHEP}
\bibliography{new.bib}
\section*{Appendix: Some Calculations for the Cumulants}
For the third term of eq. (\ref{kurtosiscalculation}), we need to calculate $\frac{d^3r_q}{dq^3}\big|_{q=1}$ and it is given by
\begin{align}
\frac{d^3r_q}{dq^3}\bigg|_{q=1}&=\frac{d}{dq}\bigg[\frac{4}{q^3}\bigg(\frac{\ell_{dS}r_q^2}{3r_q^2+\ell_{dS}^2}\bigg)-\frac{2}{q^2}\bigg(\frac{2\ell_{dS}r_q}{3r_q^2+\ell_{dS}^2}\bigg)\frac{dr_q}{dq}+\frac{12}{q^2}\bigg(\frac{\ell_{dS}r_q^2}{(3r_q^2+\ell_{dS}^2)^2}\bigg)r_q\frac{dr_q}{dq}\bigg]\bigg|_{q=1}\nonumber\\
&=-\bigg(\frac{12}{q^4}\frac{\ell_{dS}r_q^2}{3r_q^2+\ell_{dS}^2}\bigg)\bigg|_{q=1}+\bigg(\frac{12}{q^3}\frac{\ell_{dS}r_q}{3r_q^2+\ell_{dS}^2}\frac{dr_q}{dq}\bigg)\bigg|_{q=1}-\bigg(\frac{48}{q^3}\frac{\ell_{dS}r_q^3}{(3r_q^2+\ell_{dS}^2)^2}\frac{dr_q}{dq}\bigg)\bigg|_{q=1}\nonumber\\
&\;\;\;\;-\bigg(\frac{4}{q^2}\frac{\ell_{dS}}{3r_q^2+\ell_{dS}^2}\bigg(\frac{dr_q}{dq}\bigg)^2\bigg)\bigg|_{q=1}+\bigg(\frac{84}{q^2}\frac{\ell_{dS}r_q^2}{(3r_q^2+\ell_{dS}^2)^2}\bigg(\frac{dr_q}{dq}\bigg)^2\bigg)\bigg|_{q=1}\nonumber\\
&\;\;\;\;-\bigg(\frac{4}{q^2}\frac{\ell_{dS}r_q}{3r_q^2+\ell_{dS}^2}\frac{d^2r_q}{dq^2}\bigg)\bigg|_{q=1}-\bigg(\frac{24}{q^2}\frac{\ell_{dS}r_q^3}{(3r_q^2+\ell_{dS}^2)^3}6r_q\bigg(\frac{dr_q}{dq}\bigg)^2\bigg)\bigg|_{q=1}\nonumber\\
&\;\;\;\;+\bigg(\frac{12}{q^2}\frac{\ell_{dS}r_q^3}{(3r_q^2+\ell_{dS}^2)^2}\frac{d^2r_q}{dq^2}\bigg)\bigg|_{q=1}.\label{partial3rq}
\end{align}
\indent As one proceeds to the calculation of $\kappa_5^\Phi$, we have
\begin{align}
\kappa_5^{S_E}=&\frac{d^3r_q}{dq^3}\bigg(\frac{2\pi r_q}{G_Nq}\frac{dr_q}{dq}\bigg)\bigg|_{q=1}\nonumber\\
=&\frac{d^2}{dq^2}\bigg[-\frac{1}{q^2}\frac{2\pi r_q}{G_N}\frac{dr_q}{dq}+\frac{2\pi}{G_Nq}\bigg(\frac{dr_q}{dq}\bigg)^2+\frac{2\pi r_q}{G_Nq}\frac{d^2r_q}{dq^2}\bigg]\bigg|_{q=1}\nonumber\\
\equiv&\;\textbf{I}+\textbf{II}+\textbf{III},\label{kappa5}
\end{align}
where we have defined $\textbf{I}, \textbf{II}$ and $\textbf{III}$ such that
\begin{align}
\textbf{I}\equiv&-\frac{d^2}{dq^2}\bigg[\frac{1}{q^2}\frac{2\pi r_q}{G_N}\frac{dr_q}{dq}\bigg]\bigg|_{q=1}\\\nonumber
=&\bigg(\frac{12}{q^4}\frac{\pi r_q}{G_N}\frac{dr_q}{dq}\bigg)\bigg|_{q=1}-\bigg(\frac{8}{q^3}\frac{\pi }{G_N}\bigg(\frac{dr_q}{dq}\bigg)^2\bigg)\bigg|_{q=1}-\bigg(\frac{8}{q^3}\frac{\pi r_q}{G_N}\frac{d^2r_q}{dq^2}\bigg)\bigg|_{q=1}\\\nonumber
&+\bigg(\frac{6}{q^2}\frac{\pi}{G_N}\frac{dr_q}{dq}\frac{d^2r_q}{dq^2}\bigg)\bigg|_{q=1}+\bigg(\frac{2}{q^2}\frac{\pi r_q}{G_N}\frac{d^3r_q}{dq^3}\bigg)\bigg|_{q=1},
\end{align}
\begin{align}
\textbf{II}\equiv&\frac{d^2}{dq^2}\bigg[\frac{2\pi}{G_Nq}\bigg(\frac{dr_q}{dq}\bigg)^2\bigg]\bigg|_{q=1}\nonumber\\
=&\bigg(\frac{4}{q^3}\frac{\pi}{G_N}\bigg(\frac{dr_q}{dq}\bigg)^2\bigg)\bigg|_{q=1}-\bigg(\frac{8}{q^2}\frac{\pi}{G_N}\frac{dr_q}{dq}\frac{d^2r_q}{dq^2}\bigg)\bigg|_{q=1}+\bigg(\frac{4}{q}\frac{\pi}{G_N}\bigg(\frac{d^2r_q}{dq^2}\bigg)^2\bigg)\bigg|_{q=1}\\\nonumber
&+\bigg(\frac{4}{q}\frac{\pi}{G_N}\frac{dr_q}{dq}\frac{d^3r_q}{dq^3}\bigg)\bigg|_{q=1},
\end{align}
and
\begin{align}
\textbf{III}\equiv&\frac{d^2}{dq^2}\bigg[\frac{2\pi r_q}{G_N q}\frac{d^2 r_q}{dq^2}\bigg]\bigg|_{q=1}\nonumber\\
=&\bigg(\frac{4}{q^3}\frac{\pi r_q}{G_N}\frac{d^2r_q}{dq^2}\bigg)\bigg|_{q=1}-\bigg(\frac{4}{q^2}\frac{\pi}{G_N}\frac{dr_q}{dq}\frac{d^2r_q}{dq^2}\bigg)\bigg|_{q=1}-\bigg(\frac{4}{q^2}\frac{\pi r_q}{G_N}\frac{d^3r_q}{dq^3}\bigg)\bigg|_{q=1}\nonumber\\
&+\bigg(\frac{2}{q}\frac{\pi}{G_N}\bigg(\frac{d^2r_q}{dq^2}\bigg)^2\bigg)\bigg|_{q=1}+\bigg(\frac{4}{q}\frac{\pi}{G_N}\frac{dr_q}{dq}\frac{d^3r_q}{dq^3}\bigg)\bigg|_{q=1}+\bigg(\frac{2}{q}\frac{\pi r_q}{G_N}\frac{d^4r_q}{dq^4}\bigg)\bigg|_{q=1}.\label{iii}
\end{align}
Given the first, second, and third derivatives of $r_q$ at $q\rightarrow1$, the value of $\textbf{I}$ and $\textbf{II}$ can be directly determined and they give us $\textbf{I}=\frac{431}{16}\frac{\pi\ell_{dS}^2}{G_N}$ and $\textbf{II}=\frac{274}{16}\frac{\pi\ell_{dS}^2}{G_N}$. To determine the value of $\textbf{III}$ we have to find the value of $\frac{d^4r_q}{dq^4}\big|_{q=1}$, which can be calculated from
\begin{equation}
\frac{d^4r_q}{dq^4}=\frac{d^2}{dq^2}\bigg[\frac{4}{q^3}\bigg(\frac{\ell_{dS}r_q^2}{3r_q^2+\ell_{dS}^2}\bigg)-\frac{2}{q^2}\bigg(\frac{2\ell_{dS}r_q}{3r_q^2+\ell_{dS}^2}\bigg)\frac{dr_q}{dq}+\frac{12}{q^2}\bigg(\frac{\ell_{dS}r_q^2}{(3r_q^2+\ell_{dS}^2)^2}\bigg)\frac{dr_q}{dq}\bigg]\bigg|_{q=1},\label{rq4}
\end{equation}
which, again, can be separated into three parts, $\tilde{\textbf{I}}, \tilde{\textbf{II}}$, and $\tilde{\textbf{III}}$, that are given by
\begin{align}
\tilde{\textbf{I}}\equiv&\frac{d^2}{dq^2}\bigg[\frac{4}{q^3}\bigg(\frac{\ell_{dS}r_q^2}{3r_q^2+\ell_{dS}^2}\bigg)\bigg]\bigg|_{q=1}\\\nonumber
=&\bigg(\frac{48}{q^5}\frac{\ell_{dS}r_q^2}{3r_q^2+\ell_{dS}^2}\bigg)\bigg|_{q=1}-\bigg(\frac{48}{q^4}\frac{\ell_{dS}r_q}{3r_q^2+\ell_{dS}^2}\frac{dr_q}{dq}\bigg)\bigg|_{q=1}+\bigg(\frac{72}{q^4}\frac{\ell_{dS}r_q^3}{(3r_q^2+\ell_{dS}^2)^2}\frac{dr_q}{r_q}\bigg)\bigg|_{q=1}\nonumber\\
&+\bigg(\frac{8}{q^3}\frac{\ell_{dS}}{3r_q^2+\ell_{dS}^2}\bigg(\frac{dr_q}{dq}\bigg)^2\bigg)\bigg|_{q=1}+\bigg(\frac{8}{q^3}\frac{\ell_{dS}r_q}{3r_q^2+\ell_{dS}^2}\frac{d^2r_q}{dq^2}\bigg)\bigg|_{q=1}-\bigg(\frac{48}{q^3}\frac{\ell_{dS}r_q^2}{(3r_q^2+\ell_{dS}^2)^2}\bigg(\frac{dr_q}{dq}\bigg)^2\bigg)\bigg|_{q=1}\nonumber\\
=&\frac{225}{16}\ell_{dS},
\end{align}
\begin{align}
\tilde{\textbf{II}}\equiv&\frac{d^2}{dq^2}\bigg[-\frac{2}{q^2}\bigg(\frac{2\ell_{dS}r_q}{3r_q^2+\ell_{dS}^2}\bigg)\frac{dr_q}{dq}\bigg]\bigg|_{q=1}\nonumber\\
=&-\bigg(\frac{24}{q^4}\frac{\ell_{dS}r_q}{3r_q^2+\ell_{dS}^2}\frac{dr_q}{dq}\bigg)\bigg|_{q=1}+\bigg(\frac{8}{q^3}\frac{\ell_{dS}}{3r_q^2+\ell_{dS}^2}\bigg(\frac{dr_q}{dq}\bigg)^2\bigg)\bigg|_{q=1}+\bigg(\frac{8}{q^3}\frac{\ell_{dS}r_q}{3r_q^2+\ell_{dS}^2}\frac{d^2r_q}{dq^2}\bigg)\bigg|_{q=1}\nonumber\\
&-\bigg(\frac{48}{q^3}\frac{\ell_{dS}r_q^2}{(2r_q^2+\ell_{dS}^2)^2}\bigg(\frac{dr_q}{dq}\bigg)^2\bigg)\bigg|_{q=1}+\bigg(\frac{8}{q^3}\frac{\ell_{dS}}{3r_q^2+\ell_{dS}^2}\bigg(\frac{dr_q}{dq}\bigg)^2\bigg)\bigg|_{q=1}-\bigg(\frac{8}{q^2}\frac{\ell_{dS}}{3r_q^2+\ell_{dS}^2}\frac{dr_q}{dq}\frac{d^2r_q}{dq^2}\bigg)\bigg|_{q=1}\nonumber\\
&+\bigg(\frac{48}{q^2}\frac{\ell_{dS}r_q}{(3r_q^2+\ell_{dS}^2)^2}\bigg(\frac{dr_q}{dq}\bigg)^3\bigg)\bigg|_{q=1}\nonumber\\
=&\frac{493}{64}\ell_{dS},
\end{align}
and
\begin{align}
\tilde{\textbf{III}}\equiv&\frac{d^2}{dq^2}\bigg[\frac{12}{q^2}\bigg(\frac{\ell_{dS}r_q^2}{(3r_q^2+\ell_{dS}^2)^2}\frac{dr_q}{dq}\bigg)\bigg]\bigg|_{q=1}\nonumber\\
=&\bigg(\frac{72}{q^4}\frac{\ell_{dS}r_q^3}{(3r_q^2+\ell_{dS}^2)^2}\frac{dr_q}{dq}\bigg)\bigg|_{q=1}-\bigg(\frac{144}{q^3}\frac{\ell_{dS}r_q^2}{(3r_q^3+\ell_{dS}^2)^2}\bigg(\frac{dr_q}{dq}\bigg)^2\bigg)\bigg|_{q=1}-\bigg(\frac{48}{q^3}\frac{\ell_{dS}r_q^3}{(3r_q^2+\ell_{dS}^2)^2}\frac{d^2r_q}{dq^2}\bigg)\bigg|_{q=1}\nonumber\\
&+\bigg(\frac{576}{q^3}\frac{\ell_{dS}r_q^4}{(3r_q^2+\ell_{dS}^2)^3}\bigg(\frac{dr_q}{dq}\bigg)^2\bigg)\bigg|_{q=1}+\bigg(\frac{72}{q^2}\frac{\ell_{dS}r_q^2}{(3r_q^2+\ell_{dS}^2)^2}\bigg(\frac{dr_q}{dq}\bigg)^3\bigg)\bigg|_{q=1}\nonumber\\
&+\bigg(\frac{72}{q^2}\frac{\ell_{dS}r_q^2}{(3r_q^2+\ell_{dS}^2)^2}\frac{dr_q}{dq}\frac{d^2r_q}{dq^2}\bigg)\bigg|_{q=1}-\bigg(\frac{1008}{q^2}\frac{\ell_{dS}r_q^3}{(3r_q^2+\ell_{dS}^2)^3}\bigg(\frac{dr_q}{dq}\bigg)^3\bigg)\bigg|_{q=1}\nonumber\\
&+\bigg(\frac{2592}{q^2}\frac{\ell_{dS}r_q^5}{(3r_q^2+\ell_{dS}^2)^4}\bigg(\frac{dr_q}{dq}\bigg)^3\bigg)\bigg|_{q=1}-\bigg(\frac{432}{q^2}\frac{\ell_{dS}r_q^4}{(3r_q^2+\ell_{dS}^2)^3}\frac{dr_q}{dq}\frac{d^2r_q}{dq^2}\bigg)\bigg|_{q=1}\nonumber\\
&+\bigg(\frac{36}{q^2}\frac{\ell_{dS}r_q^2}{(3r_q^2+\ell_{dS}^2)^2}\frac{dr_q}{dq}\frac{d^2r_q}{dq^2}\bigg)\bigg|_{q=1}+\bigg(\frac{12}{q^2}\frac{\ell_{dS}r_q^3}{(3r_q^2+\ell_{dS}^2)^2}\frac{d^3r_q}{dq^3}\bigg)\bigg|_{q=1}\nonumber\\
=&-\frac{1503}{128}\ell_{dS}.
\end{align}
Plugging this back into eq. (\ref{rq4}) we get $\frac{d^4r_q}{dq^4}\big|_{q=1}=\frac{1283}{128}\ell_{dS}$. Thus, $\textbf{III}=\frac{3137}{64}\frac{\pi\ell_{dS}^2}{G_N}$. Plugging $\textbf{I}, \textbf{II}$, and $\textbf{III}$ back into eq. (\ref{kappa5}) give us $\kappa_5^{S_E}=\frac{5957}{64}\frac{\pi\ell_{dS}^2}{G_N}$.
\end{document}